\documentclass[final,5p,times,twocolumn]{elsarticle}
\usepackage{braket}
\usepackage{amssymb}
\usepackage{amsmath}
\usepackage{subcaption}
\usepackage[hidelinks]{hyperref}
\usepackage{natbib}
\usepackage{graphicx,color}
\usepackage{comment}
\usepackage{booktabs}   
\usepackage{multirow}   
\usepackage{makecell}   
\journal{Chaos, Solitons \& Fractals}

\begin{document}

\begin{frontmatter}

\title{Signatures of Chaos in a Quasiperiodically Driven Quantum
Impact Oscillator}

\author[a]{Deepshikha Singh}
\author[a]{Titir Mukherjee \corref{cor1}}
\ead{titirmukh96@gmail.com}
\author[a]{Soumitro Banerjee }
\cortext[cor1]{corresponding author}
\affiliation[a]{organization={Department of Physical Sciences, Indian Institute of Science Education and Research Kolkata},
            addressline={Mohanpur}, 
            city={Nadia},
            postcode={741246}, 
            state={West Bengal},
            country={India}}

\begin{abstract}

We demonstrate the emergence of quantum-chaotic dynamics in a quasiperiodically driven impact oscillator near the grazing condition. While previous studies of the quantum impact oscillator under periodic driving reported strange nonchaotic dynamics, we show that quasiperiodic driving produces robust signatures of chaos. The corresponding classical system undergoes a sharp grazing-induced transition from quasiperiodic motion to chaos, as established by bifurcation analysis, Lyapunov exponents, Fourier spectra, and the 0–1 test. In the quantum system, Shannon-entropy time series exhibit broadband spectra, 0–1 test values close to unity, and predominantly positive finite-time Lyapunov exponents for both rational and irrational driving-frequency ratios. Independent quantum diagnostics reinforce this result: the out-of-time-order correlator (OTOC) displays early-time exponential growth with nearly identical rates, while the golden-ratio drive leads to a substantially faster saturation of the OTOC. Fidelity exhibits an exponential decay regime that is approximately independent of perturbation strength. These results establish quasiperiodic driving near grazing as a mechanism for generating quantum-chaotic behavior and reveal that, although the initial onset of scrambling is largely insensitive to frequency-ratio rationality, the subsequent development of global scrambling is strongly influenced by the degree of irrationality of the drive.

\end{abstract}


\end{frontmatter}

\section{Introduction}
\label{introduction}

Classical chaotic systems governed by nonlinear differential equations exhibit a sensitivity to initial conditions, leading to seemingly unpredictable behavior even though the underlying dynamics are completely deterministic \cite{strogatz2024nonlinear, hilborn2000chaos}. Understanding how such complex behavior manifests in the quantum regime is one of the central questions in quantum chaos \cite{stockmann2007quantum, Haake1991, gutzwiller2013chaos}. 

In this paper, we consider the impact oscillator, which is known to display very rich dynamics and yet is simple enough to be amenable to formulating a quantum analog. It is a forced spring-mass system constrained by a rigid wall, allowing two types of motion: non-impacting and impacting, separated by the grazing trajectory that just touches the wall tangentially, with zero normal velocity. The grazing orbit is known to induce a variety of interesting behaviors, including grazing-induced bifurcations, multistability, and abrupt transition to chaos \citep{nordmark1991non, bernardo2008piecewise, chin1994grazing, banerjee2009invisible}. 

The rich dynamical behavior of the impact oscillator, and in particular the presence of chaos, makes it a good candidate for studying quantum chaos in the quantum domain. The impact oscillator also has a closed form of potential, which is a further advantage in the quantum setting since it allows the Hamiltonian to be constructed and diagonalized in closed form, rather than resorting to  approximations. Although the classical forced impact oscillator has been investigated by many groups because of its importance in engineering, it has not been widely studied in the quantum domain.
In this paper, we ask: Can the quantum analog of the impact oscillator display signatures of chaos?

Several approaches have been developed to identify quantum signatures of chaos. One of the earliest and most influential comes from Random Matrix Theory (RMT) \citep{bohigas1984characterization, guhr1998random}, according to which quantum systems with chaotic classical counterparts exhibit energy level statistics following the Wigner--Dyson distribution \citep{dyson1962statistical, wigner1951statistical}, while integrable systems instead follow Poisson statistics \citep{berry1977level}. Beyond RMT, widely used diagnostics include the out-of-time-order correlator (OTOC) \citep{hashimoto2017out, garcia2022out}, which measures operator growth and information scrambling and, in systems with a chaotic classical counterpart, displays an exponential rise up to the Ehrenfest time \citep{Haake1991, berman1978condition}. Another widely used measure is the fidelity \citep{peres1984stability}, which quantifies the sensitivity of quantum evolution to perturbations of the Hamiltonian; in chaotic systems, it exhibits a perturbation-independent exponential decay regime known as the Lyapunov regime \citep{jalabert2001environment, gorin2006dynamics}.

An alternative route to identifying signatures of chaos in quantum systems draws on tools from nonlinear dynamics (NLD). This requires converting the complex-valued wavefunction into a real-valued time series, which can be achieved using the expectation values of observables \citep{shankar2012principles}, the autocorrelation function \citep{nauenberg1990autocorrelation}, the $\mathcal{L}_1$-norm \citep{bernstein2018scalar}, or quantum entropies \citep{chehade2025entropy}. Common NLD tools applied to such time series include the Fourier spectrum \citep{valsakumar1997signature}, the 0--1 test \citep{gottwald2004new}, and finite-time Lyapunov exponents \citep{benettin1980lyapunov}.

Earlier work on the quantum version of the impact oscillator \citep{acharya2023signatures, acharya2026nonlinear} did not detect any evidence of the occurrence of chaos. Instead, it showed that the quantum analog of an unforced impact oscillator exhibits quasiperiodicity, and in the presence of sinusoidal forcing, it exhibits strange nonchaotic dynamics, i.e., dynamics with fractal-like structures but without exponential sensitivity to initial conditions.

Strange nonchaotic attractors (SNA) are known to occur in quasiperiodically forced systems, i.e., where the forcing function comprises two or more incommensurate frequency components \citep{prasad2001strange,feudel2006strange,LiYueGrebogi2021SNAVibroImpact,  Duan2024PiecewiseSmoothSNA}. But in the quantum analog of the impact oscillator, such behavior arose in a sinusoidally forced system. This observation naturally raises the question: What behavior would such a system exhibit under quasiperiodic forcing? Can chaos arise in such a system? This is the question that this paper addresses. 

In addition, we explore the effect of the  rationality of the frequency ratio $\omega_r = \omega_1/\omega_2$. A rational value of $\omega_r$ effectively corresponds to a periodic drive, whereas irrational values produce quasiperiodic forcing. The golden ratio $\omega_r = (\sqrt{5} + 1) / 2$ represents the most irrational case in the sense of the continued-fraction theory \cite{khinchin1964continued}, being maximally resistant to a rational approximation. In this work, we ask: how does the rationality of $\omega_r$ shape the signatures of chaos of a quantum impact oscillator near the grazing condition? 

Although the classical impact oscillator is a well-explored system, such a system under quasiperiodic drive has not been explored to date. So, we study the quasiperiodically driven impact oscillator in both its classical and quantum realizations, focusing on the dynamics near
the grazing condition. For the classical system, we find a sharp transition from quasiperiodic motion to chaos at grazing, confirmed consistently by the bifurcation diagram, the Lyapunov exponents, the
Fourier spectrum, and the 0--1 test. 
We propagate the quantum system's wavefunction using a fourth-order commutator-free exponential time-propagator (CFET). We use the Shannon entropy as a real-valued time series representing the dynamics and apply the diagnostic tools of nonlinear dynamics. We find that the system has a broadband Fourier spectrum, the 0--1 test returns values close to unity, and the
finite-time Lyapunov exponents are predominantly positive for every driving frequency ratio we consider. These observations show that quasiperiodic driving near grazing pushes the system into chaotic behavior. The out-of-time-order correlator and fidelity, used as independent quantum-mechanical diagnostics, corroborate this picture and further show that the degree of irrationality of
$\omega_r$ governs how quickly chaos develops.

The paper is organized as follows. In Section~\ref{Methodology}, we introduce the classical and quantum impact oscillator under quasiperiodic drive, together with the numerical scheme used to propagate the wavefunction. In Section~\ref{cl_F}, we analyze the classical dynamics and show that the system undergoes a sharp transition from quasiperiodic motion to chaos at grazing. In Section~\ref{qfo}, we analyze the quantum dynamics and show that quasiperiodic driving pushes the system into chaotic behavior near grazing, in contrast to strange nonchaotic dynamics reported under periodic driving. Section~\ref{summary} summarizes the findings and outlines future directions.

\section{Methodology}
\label{Methodology}

\subsection{The Classical Impact Oscillator}

\begin{figure}[h!]
    \centering
    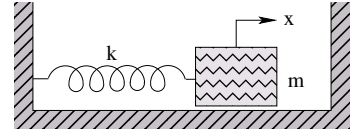
    \caption{Schematic diagram of a simple impact oscillator}
    \label{spring mass}
\end{figure}

A simple impact oscillator is a spring-mass system constrained by a hard wall (as shown in Fig.~\ref{spring mass}). 
The parameters involved in the system are: 
the spring constant $k$, the mass of the particle $m$, and the wall position $x_w$. 

\begin{figure}[h!]
    \centering
    \includegraphics[width= 0.8\linewidth]{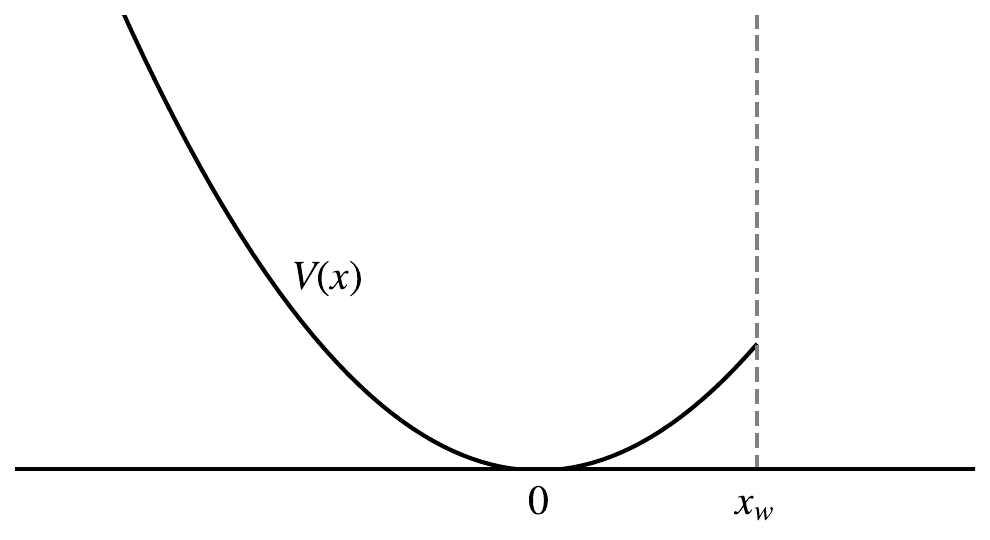}
    \caption{The graphical representation of the impact oscillator potential $V(x)$.}
    \label{V(x)}
\end{figure}

The potential of a simple impact oscillator can be described by
\begin{equation}
V(x)=
\begin{cases}
\frac{1}{2}kx^2 & \text{if } x < x_w \\
\infty & \text{if } x \geq x_w .
\end{cases}
\label{eq1}
\end{equation}
 The graphical representation of the potential function described in Eq.~(\ref{eq1}) is shown in Fig.~\ref{V(x)}.

\noindent The system is driven by two frequency components, $\omega_1$ and $\omega_2$, with amplitudes $a$  and $b$, respectively.
The potential function of such a forced impact oscillator is
\begin{equation}
    V(x,t)  =  V(x)  +  x\ F(t),
    \label{eq2}
\end{equation}
where $F(t)$ is the externally applied force, which is of the form
\begin{equation}
    F(t) = a \sin(\omega_1 t) + b \cos(\omega_2  t), \label{forcing}
\end{equation}

Between impacts, the system's dynamics are equivalent to those of a driven harmonic oscillator. Therefore, we use the analytic solution (\ref{eqm}) of the classical equation of motion between impacts
\begin{equation}
    \dot x = v, \quad \dot v = -\frac{k}{m}x + \frac{F(t)}{m}, \label{eqm}
\end{equation}
where $m$ is the mass of the particle. Further, at each impact, the velocity is reversed, assuming perfectly elastic collisions without dissipation.

\subsection{Quantum Dynamical Framework}

The quantum analog of the impact oscillator is described by the evolution of a wavefunction $\psi(x)$, in the potential described in Eq.~(\ref{eq2}). The dynamics of the quantum system is governed by the time-dependent Schr\"odinger equation
\begin{equation}
    i\hbar\ \frac{\partial \psi (x,t)}{\partial t} = \mathcal{H}\ \psi(x,t) =  \left(-\frac{\hbar^2}{2m} \frac{d^2}{dx^2}+ V(x,t)\right) \psi(x,t)
    \label{3}
\end{equation}

The Hamiltonian $\mathcal{H}$ acts on square-integrable wavefunctions defined on the half-line $(-\infty, x_w]$, subject to a Dirichlet condition at the wall — the wavefunction vanishes at $x = x_w$, reflecting the infinite potential barrier there — together with the usual requirement that the wavefunction and its first derivative vanish as $x \to -\infty$, so that the state remains normalisable. $V(x,t)$ is the potential of the system described by (\ref{eq2}). We can also write the Hamiltonian $\mathcal{H} = H_0 + V(t)$, where $H_0$ is the static (time-independent) part of the Hamiltonian and $V(t)$ is the time-dependent part.

Since closed-form solutions of the quasiperiodically driven quantum impact oscillator do not exist, we evolve the system numerically using the energy basis of the static Hamiltonian ($H_0$). We computed the eigenvalues $E_n$ and the corresponding eigenstates $\ket{\phi_n}$  using the Numerov-Cooley method \citep{numerov1933publs, cooley1961improved} and verified the results by diagonalizing the Hamiltonian matrix \citep{izaac2018computational}.

\subsection{Numerical Simulation}

\label{sec:numerics}
To numerically solve the Schr\"odinger equation (\ref{3}) with a time-dependent Hamiltonian, we use commutator-free exponential time-propagators (CFETs) \cite{alvermann2011high, Alvermann_2012}. It is motivated by the Magnus expansion, but avoids the use of commutators.

We can write Eq.~(\ref{3}) in the form,

\begin{equation}
   \frac{d}{dt} \psi(t) = A(t)\psi(t), \label{5}
\end{equation}
where $A(t) = -i\mathcal{H}/\hbar$. We construct the propagator such that $\psi(t_2) = U(t_2,t_1)\,\psi(t_1)$, which also satisfies the initial value problem,
\begin{equation}
    \frac{d}{dt_2} {U}(t_2,t_1) = A(t_2)\ {U}(t_2,t_1),\quad
    {U}(t_1,t_1) = \mathcal{I}, \label{6}
\end{equation}
where $\mathcal{I}$ is the identity operator and $t_1$ is the initial time. We set $t_1 = t_0 = 0$; to obtain the propagator starting from an arbitrary initial time $t_0$, one can always perform the variable substitution $t \to t + t_0$.

Using the Magnus expansion \citep{Alvermann_2012, blanes2009magnus}, we can write
\begin{equation}
   {U}(t, 0) = e^{\Omega(t)}, \quad \Omega(t) = \sum_{n=1}^{\infty}\Omega_n(t). \label{7}
\end{equation}
The Magnus expansion represents $\Omega(t)$ in (\ref{7}) as a series expansion, where $\Omega_n$ is an $n$-fold integral of $(n-1)$-fold commutators of $A(t)$.
The first two terms of the Magnus expansion are given by,

\begin{equation}
\Omega_1(t)=\int_0^t A(t_1)\,dt_1,
\end{equation}

\begin{equation}
\Omega_2(t)=\frac{1}{2}\int_0^t dt_1\int_0^{t_1}dt_2
[A(t_1),A(t_2)].
\end{equation}

CFET is used to avoid nested commutators, which make direct implementation of the Magnus expansion computationally expensive for higher-order propagation. The propagator is approximated as a product of
exponentials of weighted Hamiltonians evaluated at intermediate times
within a single time step.

\begin{equation}
\tilde{U}_{\mathrm{CFET}}(\delta t)
=
e^{\;\chi_1}
e^{\;\chi_2}
\cdots
e^{\;\chi_s},
\end{equation}
where each exponential operator is expressed as a weighted sum of the operator $A(t)$ evaluated at different quadrature points,
\begin{equation}
\chi_i
=
\delta t
\sum_{m=1}^{M}
g_{i,m}\;A(x_m\delta t).
\end{equation}

The number of exponentials $s$ and the number of quadrature points $M$ are not chosen freely; rather, they are the smallest values that allow the propagator to match the Magnus expansion up to the desired order $N$ \citep{alvermann2011high, Alvermann_2012}. In other words, more exponentials and quadrature points are needed to reach higher accuracy, but the scheme always uses as few as possible to keep the computation efficient. In practice, $x_m$ are taken to be the Gauss--Legendre
quadrature points, since this choice yields the required accuracy for
the time-ordered integrals with the evaluation points $M$.

In the present work, we employ the optimized fourth-order CFET. The corresponding propagator is written as \citep{Alvermann_2012}
\begin{align}
U^{(4)}_{\mathrm{cf}}(\delta t)
&= \exp\left[\delta t \left\{ g_1 A(x_1 \delta t) + g_2 A(x_2 \delta t)  + g_3 A(x_3 \delta t) \right\}\right] \nonumber \\
& \times \exp\left[\delta t \left\{ g_4 A(x_1 \delta t)  + g_5 A(x_2 \delta t) + g_4 A(x_3 \delta t)  \right\}\right] \nonumber \\
& \times \exp\left[\delta t \left\{ g_3 A(x_1 \delta t)  + g_2 A(x_2 \delta t)  + g_1 A(x_3 \delta t)  \right\}\right]
\end{align}

\noindent The Gauss–Legendre quadrature points $x_m$ are
\begin{equation}
     x_1 = \frac{1}{2} - \frac{\sqrt{3}}{20}, \quad
 x_2 = \frac{1}{2},\quad  
 x_3 = \frac{1}{2} + \frac{\sqrt{3}}{20}
\end{equation}
and the coefficients $g_{i,m}$ (obtained by comparison with the Magnus
expansion) are given by,
\begin{equation}
g_1 = \frac{37}{240} - \frac{10}{87}\sqrt{\frac{5}{3}}, \quad
g_2 = -\frac{1}{30}, 
\end{equation}
\[
g_3 = \frac{37}{240} + \frac{10}{87}\sqrt{\frac{5}{3}},\quad
g_4 = -\frac{11}{360}, \quad
g_5 = \frac{23}{45}.
\]

\section{Dynamics of the Classical Forced Impact Oscillator}
\label{cl_F}

\begin{figure*}[t]
    \centering
    \includegraphics[width=\textwidth]{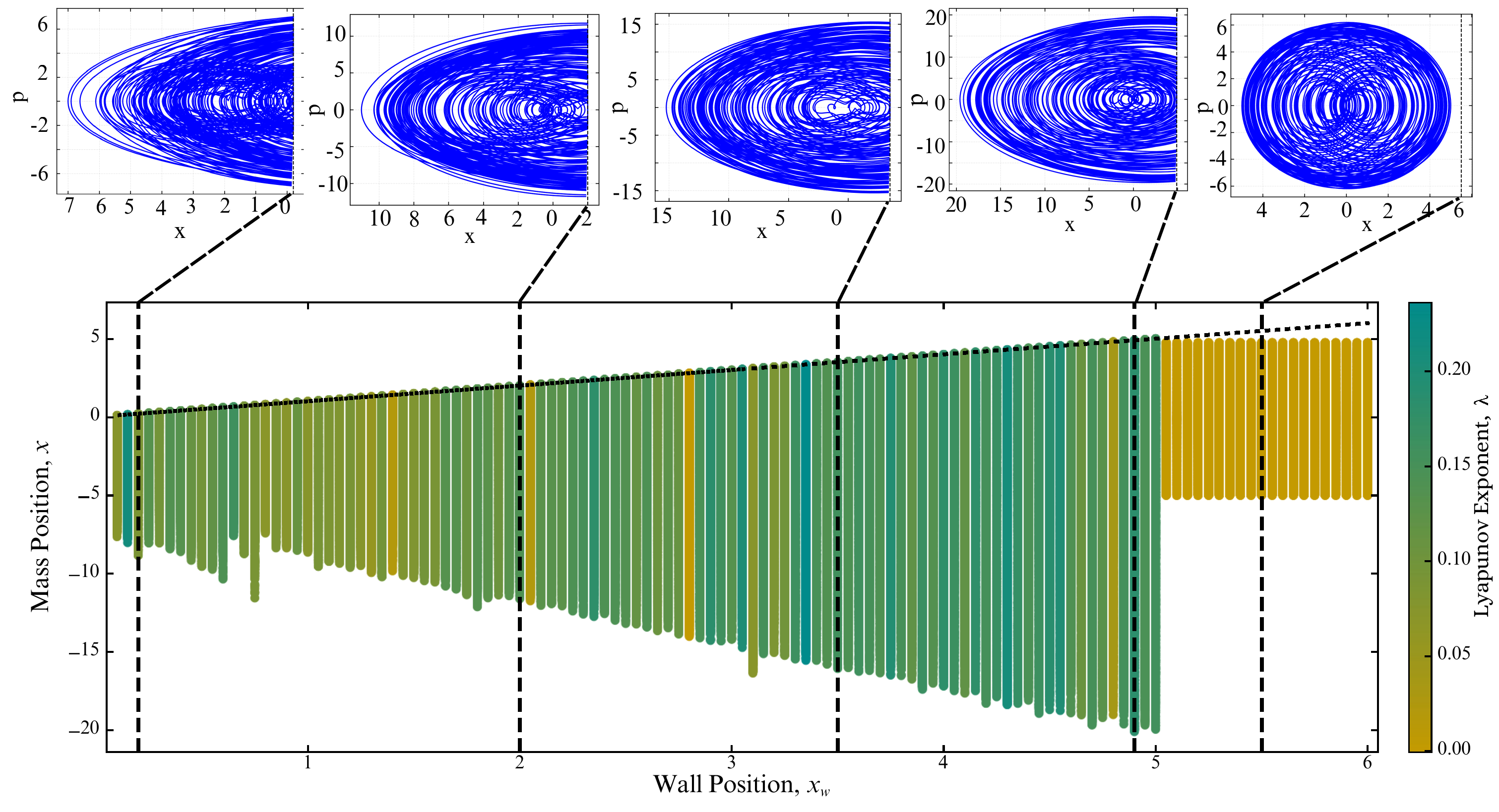}
    \caption{Bifurcation diagram of the impact oscillator system with quasiperiodic drive. Here, the color represents the Lyapunov exponent for each wall position. The panels on the top are the phase-space trajectories for selected wall positions.}
    \label{fig:bif}
\end{figure*}

We study the classical dynamics of the quasiperiodic driven impact oscillator. 

We take the parameter values $k = 1$, $m = 1$, and as a result the natural frequency of the system turns out to be $\omega_0 = 1$. The forcing consists of two incommensurate frequencies $$\omega_1 = \frac{\sqrt{5}+1}{\sqrt{2}} \quad \mbox{and} \quad \omega_2 = \sqrt{2}$$ whose ratio corresponds to the golden ratio, with amplitudes $a = 0.606$ and $b = 2.421$.  For the given parameter values, the grazing condition occurs at $x_w = 5.0$. The position of the confining boundary $x_w$ is varied to study its effect on the system.


\begin{figure}
    \centering

    \begin{subfigure}[b]{\linewidth}
        \centering
        \includegraphics[width=0.76\linewidth]{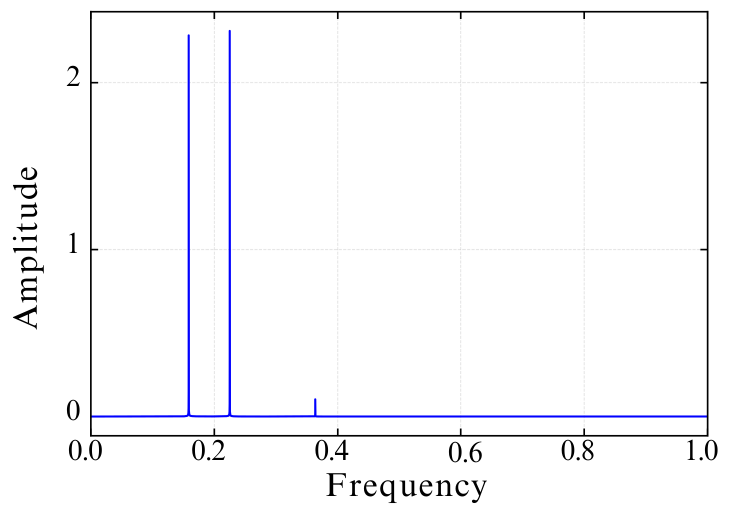}
        \caption{}
        \label{fig:a}
    \end{subfigure}

    \vspace{0.3cm}

    \begin{subfigure}[b]{\linewidth}
        \centering
        \includegraphics[width=0.76\linewidth]{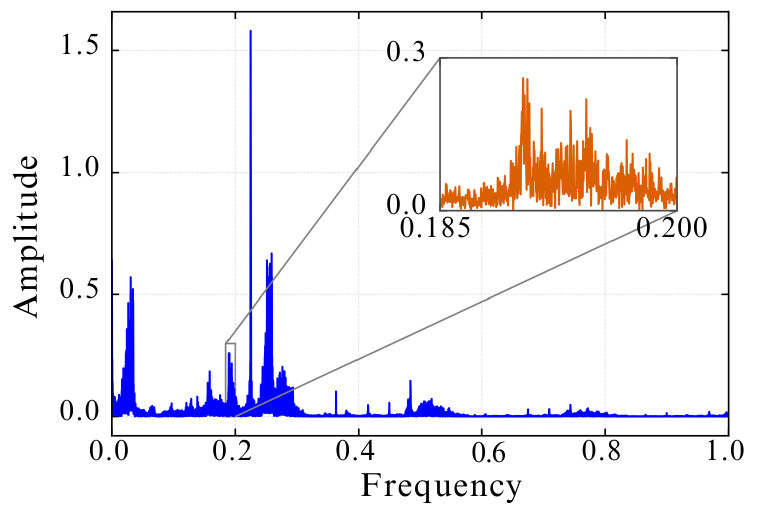}
        \caption{}
        \label{fig:b}
    \end{subfigure}

    \caption{Fourier spectrum of the position $x$ time series. (a) $x_w = 5.5$, shows three discrete peaks corresponding to the three frequencies $\omega_i/2\pi$, where $i = 0,1,2$. (b) $x_w = 4.9$ shows a continuous spectrum, clearly seen in the inset.}
    \label{fig:combined}
\end{figure}

\begin{figure}[tb]
    \centering
    \includegraphics[width=0.9\linewidth]{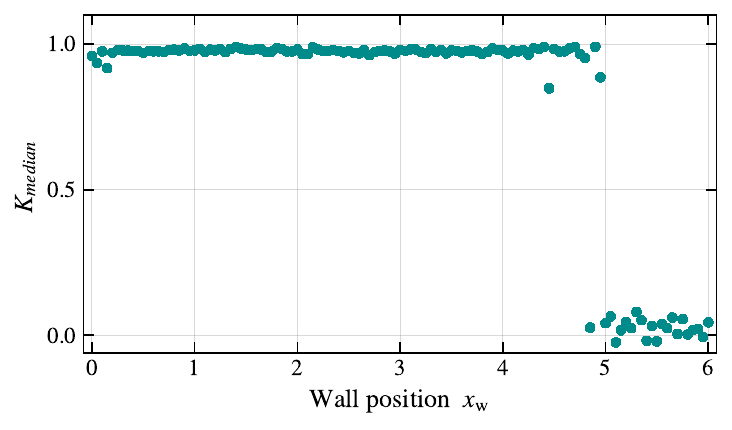}
    \caption{The result of the 0-1 test with varying wall position $x_w$. $K_{median} \approx 1$ suggests chaotic dynamics, and $K_{median} \approx 0$ indicate a regular dynamics.}
    \label{cl_01test}
\end{figure}

\begin{figure*}[tb]
    \centering
    \includegraphics[width=0.9\linewidth]{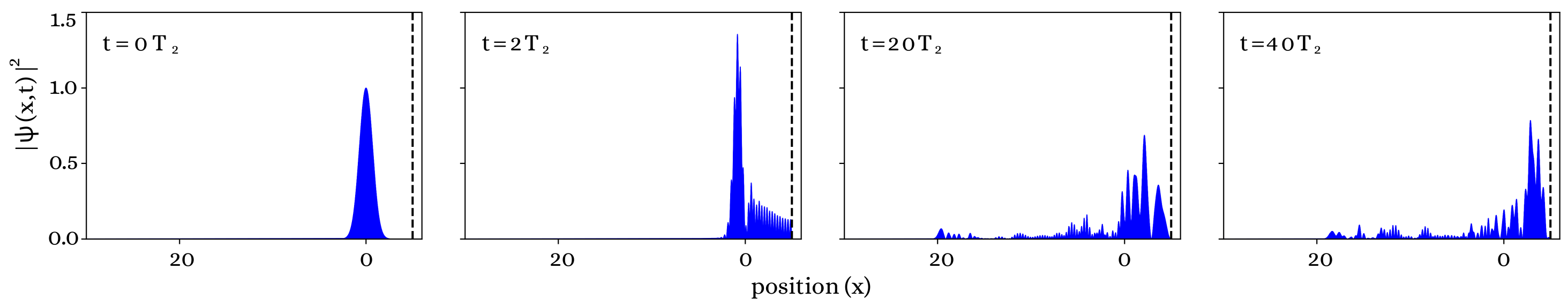}
    \caption{Snapshots of the probability density at four different times 
    showing the temporal evolution of the wavefunction under quasiperiodic 
    driving. The dashed vertical line marks the wall position, $x_w = 5$. 
    The driving frequencies are chosen as $\omega_1 = \frac{\sqrt{5}+1}
    {\sqrt{2}}$ and $\omega_2 = \sqrt{2}$, while the grazing condition is 
    observed for $a=0.606$ and $b=2.421$. Here $T_2 = 2\pi/\omega_2$.}
    \label{prob_den}
\end{figure*}

Figure~\ref{fig:bif} shows the bifurcation diagram of the system obtained by 
varying the wall position $x_w$, together with the corresponding Lyapunov exponent $\lambda$ shown as the color bar. The five panels on top show the phase-space trajectories at selected wall positions, marked by the dashed lines connecting them to the bifurcation diagram below. For constructing the bifurcation diagram, we used a stroboscopic map, we sample the trajectory at intervals $T_1 = 2\pi/\omega_1$. The largest Lyapunov exponent is calculated using the Benettin algorithm \citep{benettin1980lyapunov, bezanson2017julia, datseris2018dynamicalsystems}. 

We find that the system exhibits a sudden transition from a quasiperiodic orbit to a large-amplitude chaotic orbit following the grazing event that occurs at $x_w=5$. The Lyapunov exponents corroborate this observation, with the color bar indicating a shift from $\lambda = 0$ to positive values across the grazing boundary. This grazing-induced transition from a regular behavior to chaos is a typical attribute of the impact oscillator \citep{nordmark1991non, ing2008experimental, ING2010312}.

The Fourier spectrum (computed using position $x$ time series) shown in Fig.~\ref{fig:combined} clearly distinguishes between the two different dynamical regimes. Before grazing, i.e., for $x_w > 5$ in Fig.~\ref{fig:a}, three distinct peaks can be seen which correspond to the three frequencies in the system $\omega_0$, $\omega_1$ and $\omega_2$. This suggests that the system shows quasiperiodic dynamics. After grazing (Fig.~\ref{fig:b}), i.e., for $x_w < 5$, a continuous spectrum can be seen (see inset), which suggests that the system exhibit chaotic dynamics \citep{huberman2017power}.

A consistent trend from regular to chaotic dynamics is also captured by the 0–1 test Fig.~\ref{cl_01test}. The 0–1 test operates on a sampled time series and produces a scalar output $K_{‌\rm median}$, where values of $K_{\rm median} \approx 0$  are indicative of periodic or quasiperiodic behavior, while values $K_{\rm median} \approx 1$ signify the presence of chaotic dynamics \cite{gottwald2004new}. The test statistic $K_{\rm median}$ for the system changes sharply from near zero to near unity at the grazing point. Collectively, these results confirm that the system undergoes quasiperiodic motion prior to grazing, followed by a transition to chaotic dynamics immediately thereafter.

\section{Dynamics of the Quantum System}
\label{qfo}

Having characterized the grazing condition and its dynamical signatures in the classical system, we now turn to its quantum-mechanical counterpart. In the classical case, grazing is a sharply defined event. Due to the spatially extended nature of the wave function, the grazing condition cannot be defined in the same manner as in the classical system. To establish an analogous grazing condition in the quantum case, we identify the grazing condition as the set of parameters for which the peak of the probability density $|\Psi(x,t)|^2$ of the wavefunction evolving under a harmonic potential (i.e., in the absence of the wall) just reaches the position of the wall.

For the numerical simulations, we take the initial condition as a Gaussian 
wave packet,
\begin{equation}
\psi(x,t=0) = \frac{1}{(2\pi)^{1/4}\sqrt{\sigma}} 
\exp\left[-\left(\frac{x - x_0}{2\sigma}\right)^2\right] 
\label{wavefunction}
\end{equation}
of coherent-state width, with mean position $x_0 = 0$ and variance $\sigma^2 = \frac{\hbar}{2\sqrt{km}}$. 

We work in natural units with $k = m = \hbar = 1$. In these units 
$\sigma^2 = 1/2$, the initial coherent state is well within the 
quantum regime. The classical limit corresponds to $\hbar_{\mathrm{eff}} 
\to 0$, which is not explored here. The wavefunction is evolved under the same potential described in 
(\ref{eq2}). Fig.~\ref{prob_den} shows the probability density 
profiles evolving over time. It is clear that the wavefunction spreads out after interacting with the wall, suggesting complex underlying dynamics. The parameter values are, $\omega_1 = \frac{\sqrt{5}+1}{\sqrt{2}}$, $\omega_2 = \sqrt{2}$, $a=0.606$, $b=2.421$ and $x_w = 5$. Time is in units of $T_2 = 2\pi/\omega_2$. 

\subsection{Results}

We investigate the dynamics of the forced quantum 
impact oscillator using both quantum-mechanical and nonlinear dynamical 
(NLD) measures. Our primary objective is to examine how the behavior of the system
 changes as the driving frequencies $\omega_1$ and $\omega_2$ 
are tuned such that their ratio, $\omega_r = \omega_1/\omega_2$, changes 
from rational to irrational values, thereby transitioning from periodic 
to quasiperiodic forcing. The values of $\omega_r$ with which we have worked 
in this paper are $\omega_r = 2, 1.6$ and $(\sqrt{5}+1)/2$. The dynamics of 
the system is then investigated in the vicinity of the grazing condition 
to explore the emergence of novel dynamical behavior arising from this 
frequency modulation.

\subsubsection{Fourier Spectrum}
\label{subsec:fft01}
\begin{figure*}[t]
    \centering
    \includegraphics[width=0.9\linewidth]{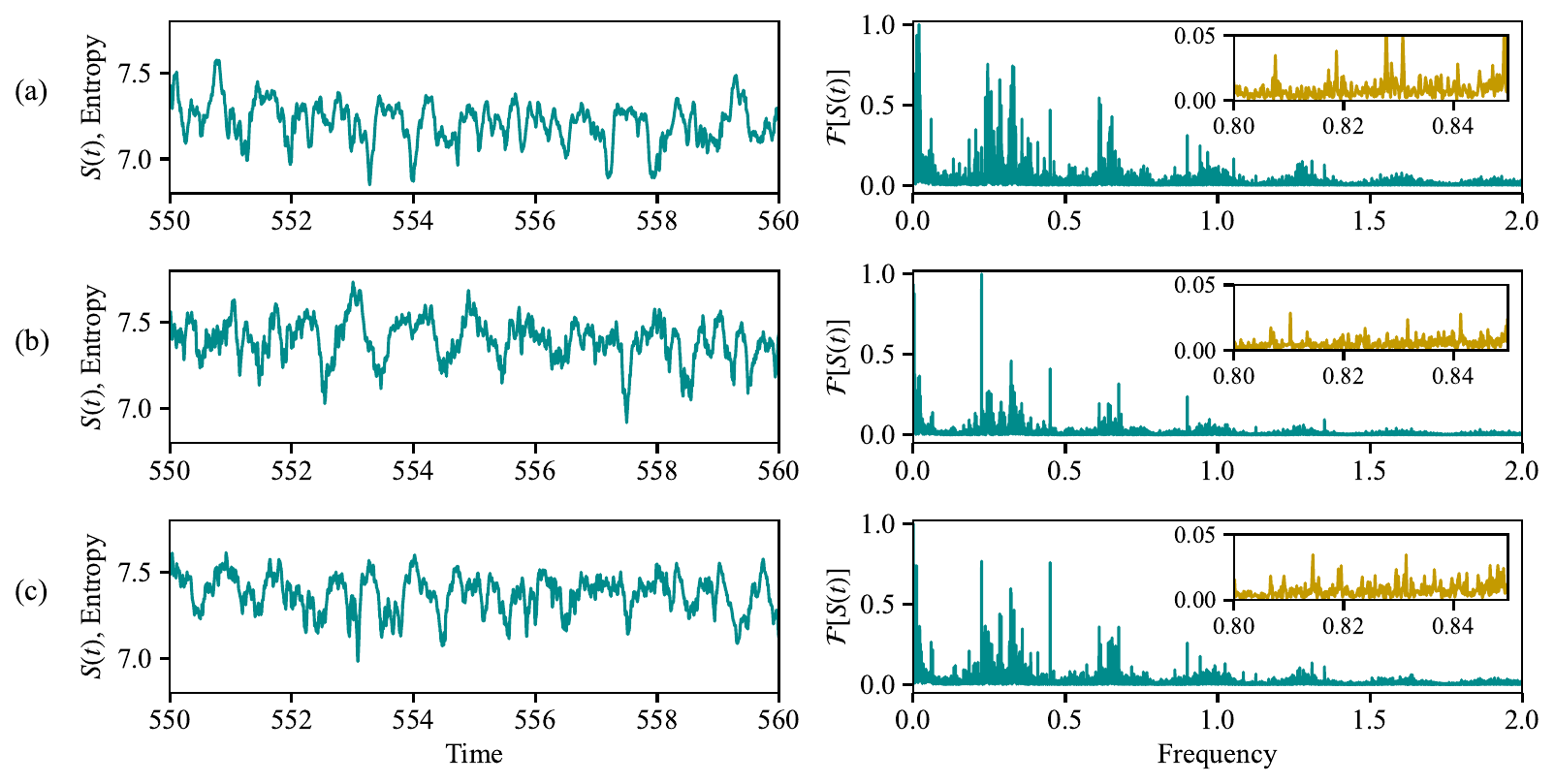}
    \caption{Time evolution of the entropy $S(t)$ (left column) and the 
    corresponding normalized Fourier spectra $\mathcal{F}[S(t)]$ (right 
    column) for three forcing-frequency ratios: (a) $\omega_r = 2$, 
    (b) $\omega_r = 1.6$, and (c) $\omega_r = (\sqrt{5}+1)/2$. The Fourier 
    spectra reveal the frequency components of the entropy, while the insets 
    provide a magnified view of the spectral region in the frequency interval $0.8 \leq f \leq 0.85$. Time is in the units of $T_2 = 2\pi/\omega_2$.}
    \label{entrpy_fft}
\end{figure*}

\begin{figure*}[tb]
    \centering
   
        \centering
        \includegraphics[width=\linewidth]{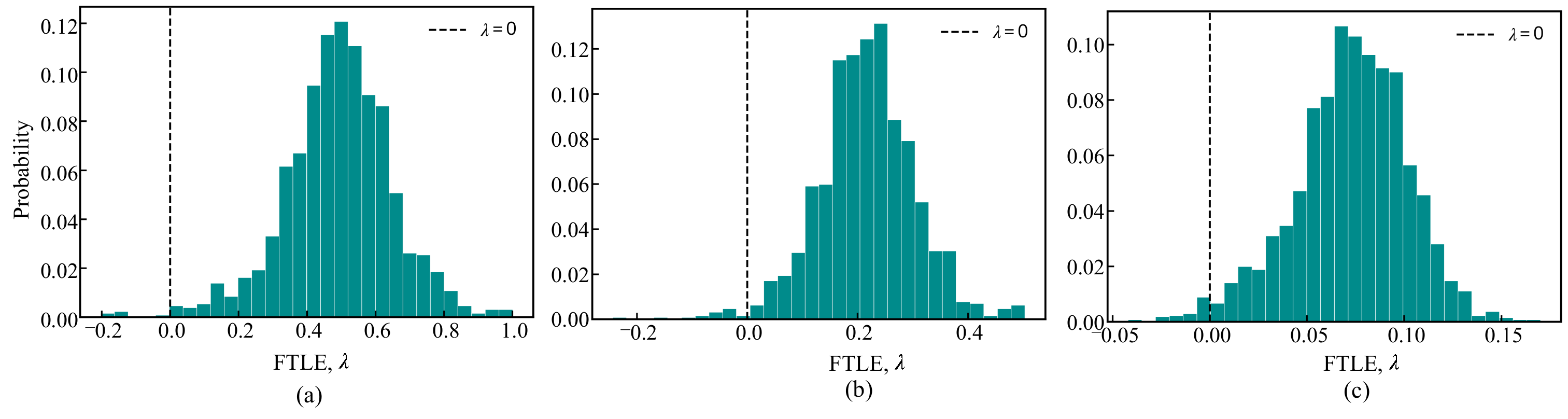}
        \label{ftle:a}

    \caption{Probability distribution of finite-time Lyapunov exponents 
    (FTLEs) computed from the entropy time series for (a) $\omega_r = 2$, 
    (b) $\omega_r = 1.6$, and (c) $\omega_r = (\sqrt{5}+1)/2$. The dashed 
    vertical line marks $\lambda = 0$, separating the chaotic 
    $(\lambda > 0)$ and regular $(\lambda < 0)$ regions.}
    \label{ftle}
\end{figure*}

\begin{figure*}[h!]
\centering
    \begin{subfigure}[t]{0.49\textwidth}
        \centering
        \includegraphics[width=\linewidth]{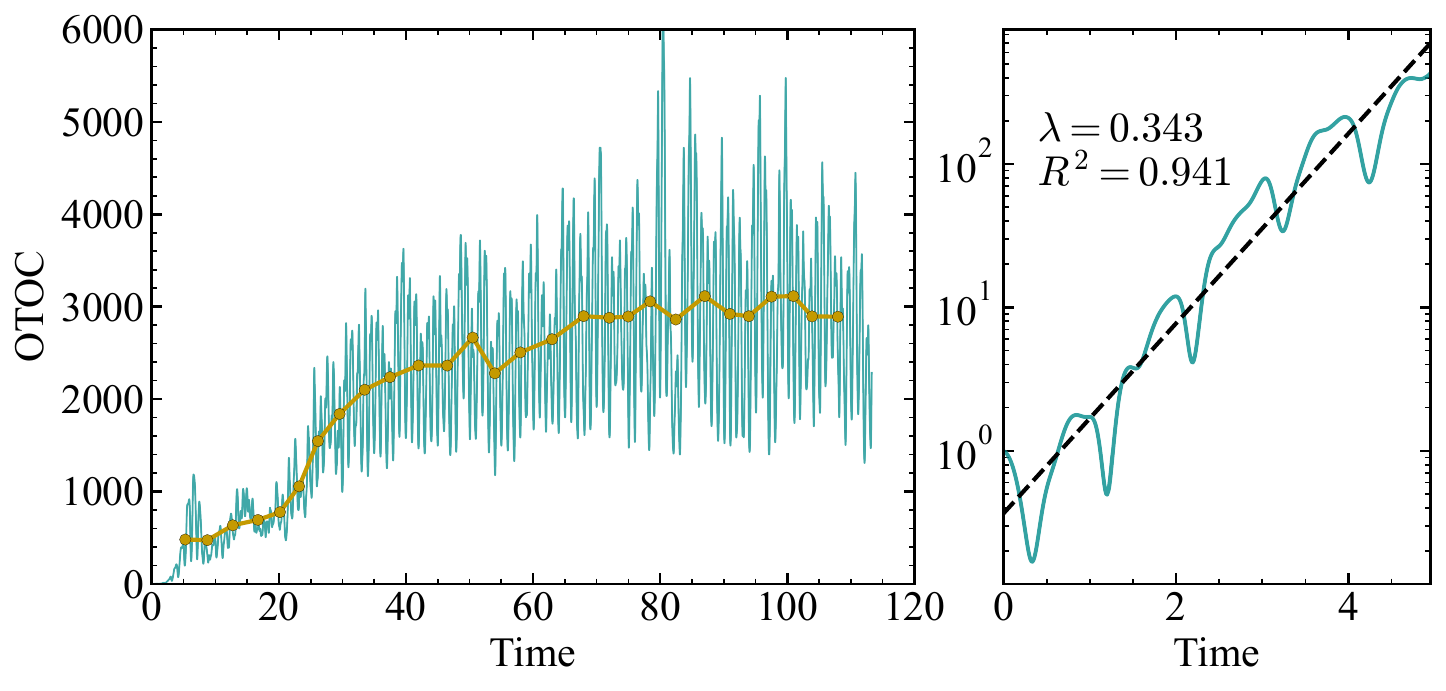}
        \caption{$\omega_r = 2$}
        \label{fig:OTOC_2}
    \end{subfigure}
    \hfill
    \begin{subfigure}[t]{0.49\textwidth}
        \centering
        \includegraphics[width=\linewidth]{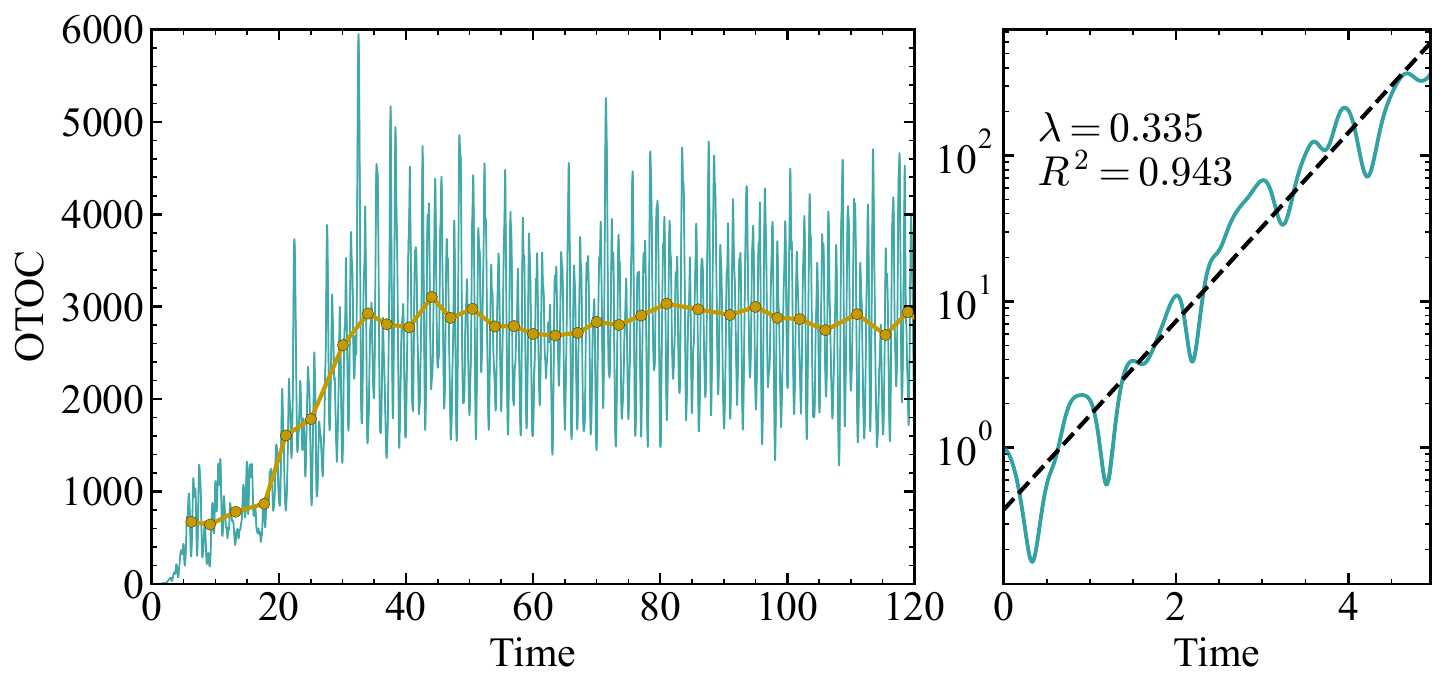}
        \caption{$\omega_r = 1.6$}
        \label{fig:OTOC_1.6}
    \end{subfigure}

    \vspace{0.8em}

    \begin{subfigure}[t]{0.49\textwidth}
        \centering
        \includegraphics[width=\linewidth]{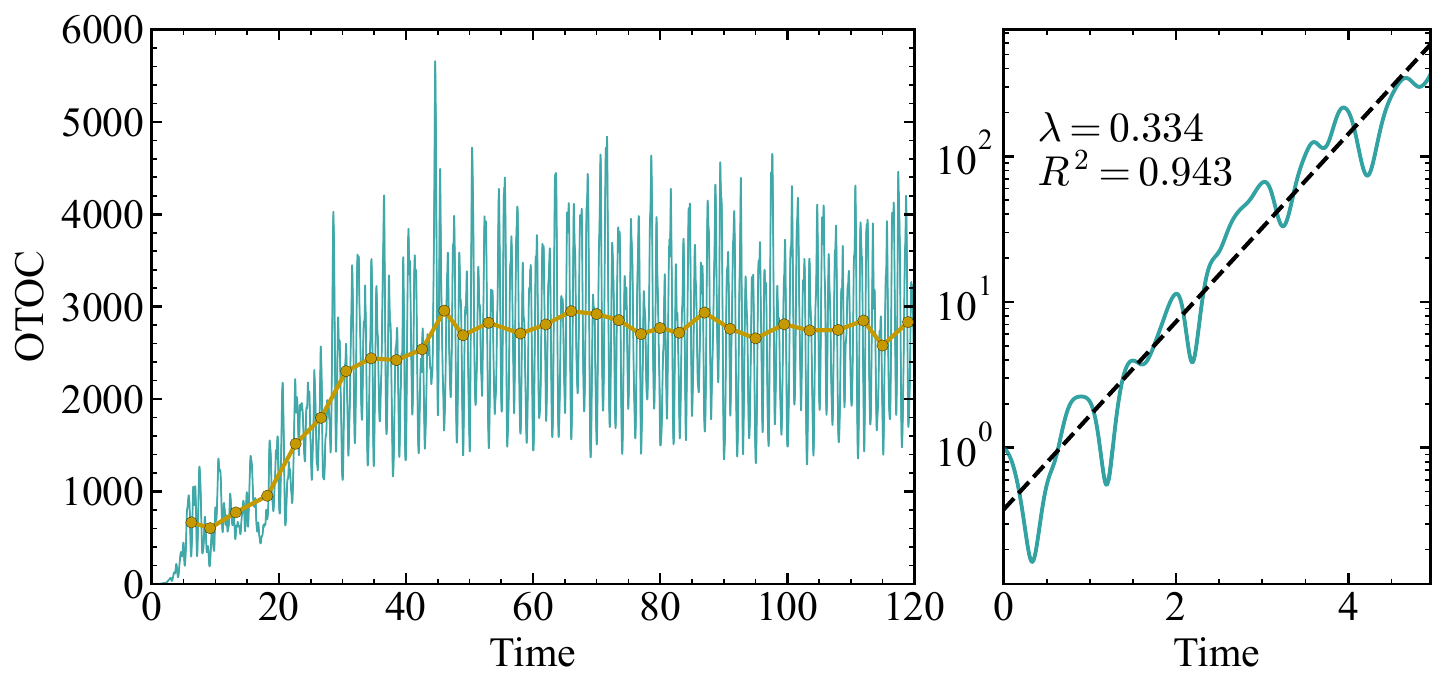}
        \caption{$\omega_r = (\sqrt{5}+1)/2$}
        \label{fig:OTOC_golden}
    \end{subfigure}

\caption{OTOC as a function of time for different forcing-frequency ratios
$\omega_r = \omega_1/\omega_2$. Here, time is in units of $T_2 = 2\pi/\omega_2$, where $\omega_2 = \sqrt{2}$ is held fixed while
$\omega_1$ is varied to obtain the different values of $\omega_r$ considered.
In each panel, the left plot shows the full time evolution of the OTOC
(blue-green) together with its time-averaged trend (yellow), while the right
plot shows the corresponding early-time growth on a semi-logarithmic scale,
along with an exponential fit $\propto e^{\lambda t}$ (dashed line); the
extracted growth rate $\lambda$ and the coefficient of determination $R^2$
are annotated directly in this panel. (a) $\omega_r = 2$, with
$\lambda = 0.343$ and $R^2 = 0.941$. (b) $\omega_r = 1.6$, with
$\lambda = 0.335$ and $R^2 = 0.943$. (c) $\omega_r = (\sqrt{5}+1)/2$ (golden
ratio), with $\lambda = 0.334$ and $R^2 = 0.943$.}
\label{fig:OTOC_all}
\end{figure*}
In order to apply the tools of nonlinear dynamics, we require a real-valued time series from the 
complex-valued wave function. We compute the Shannon entropy of the probability density of the wave function in the position basis:
\begin{equation}
S(t) = - \int |\Psi(x,t)|^2 \, \log\!\left(|\Psi(x,t)|^2\right) \, dx
\label{5.5}
\end{equation}

The Shannon entropy given in (\ref{5.5}) is related to the uncertainty in the 
position measurements and characterizes the spread of the wave function 
\cite{shannon1948mathematical}. This real-valued time series $S(t)$ is then 
used as the input for all subsequent diagnostics. The length of the time series is taken to be $3000\;T_2$, where
$T_2 = 2\pi/\omega_2$.

The entropy time series for the time window $550T_2$ to $560T_2$ is plotted
in the left panel of Fig.~\ref{entrpy_fft}. We compute the Fourier
spectrum of the entropy time series, shown in the right panel of
Fig.~\ref{entrpy_fft}. For (a) $\omega_r = 2$, (b) $\omega_r = 1.6$, and
(c) $\omega_r = (\sqrt{5}+1)/2$, the Fourier spectrum exhibits a broadband
continuous structure, in contrast to the discrete spectral peaks
expected for regular or quasiperiodic dynamics. This provides supporting evidence for chaotic dynamics \cite{huberman2017power}.

\begin{table}[b]
\centering
\caption{Values of $K_{\mathrm{median}}$ obtained from the 0--1 test 
for different forcing-frequency ratios $\omega_r$.}
\label{tab:01quantum}
\begin{tabular}{cc}
\hline\hline
$\omega_r$ & $K_{\mathrm{median}}$ \\
\hline
$1.6$           & $0.84$ \\
$({\sqrt{5}+1})/{2}$   & $0.91$ \\
$2$             & $0.73$ \\
\hline\hline
\end{tabular}
\end{table}

\subsubsection{The 0--1 Test}

To further quantify the nature of the dynamics, we apply the 0--1 test 
\cite{gottwald2004new} to the entropy time series. The 0--1 test operates 
on a scalar time series and produces a diagnostic value $K_{\mathrm{median}}$, where $K_{\mathrm{median}} \approx 1$ indicates chaotic dynamics and 
$K_{\mathrm{median}} \approx 0$ indicates regular or quasiperiodic 
behavior. The results are listed in Table~\ref{tab:01quantum}. 
All three frequency ratios yield $K_{\mathrm{median}}$ values 
close to one, confirming chaotic dynamics. The strongest chaotic
signature is observed for the golden-ratio driving $\omega_r =(\sqrt{5}+1)/2$, which gives $K_{\mathrm{median}} = 0.91$.

\subsubsection{Finite-Time Lyapunov Exponents}
\label{subsec:ftle}

To further characterize the dynamical nature of the system, we compute the 
finite-time Lyapunov exponents (FTLEs) \cite{grassberger1988scaling, 
kapitaniak1995distribution} from the entropy time series $S(t)$. The FTLE 
quantifies the average exponential rate of divergence of nearby trajectories 
over a finite time window. A positive FTLE indicates local exponential 
separation of trajectories, while a negative value corresponds to local 
convergence.

Fig.~\ref{ftle} shows the probability distribution of FTLEs for $\omega_r = $ (a) $ 2$, (b) $1.6$, and (c) $(\sqrt{5}+1)/2$. In all three cases, the distribution is 
strongly concentrated in the positive region ($\lambda > 0$), with only a 
small negative tail attributable to finite-time sampling effects. This predominantly positive FTLE distribution is a well-established 
characteristic signature of chaotic dynamics \cite{prasad1999characteristic} 
and is consistent with the Fourier and 0--1 test results.

\subsubsection{Out-of-Time-Order Correlator}
\label{subsec:otoc}

To complement the nonlinear dynamical analysis with a quantum-mechanical diagnostic, we 
compute the out-of-time-order correlator (OTOC), which provides a direct 
measure of operator growth and information scrambling in the quantum system. 
The OTOC is defined as

\begin{equation}
C_T \equiv - \langle[W(t), V(0)]^2\rangle,
\label{eq2.7}
\end{equation}
where $\langle\cdots\rangle$ denotes the thermal average 
\cite{hashimoto2017out}:
\begin{equation}
\langle \mathcal{O} \rangle =
\frac{\mathrm{Tr}\left[e^{-\beta \mathcal{H}}\mathcal{O}\right]}
{\mathrm{Tr}\left[e^{-\beta \mathcal{H}}\right]},
\end{equation}
with $\beta = 1/(k_B T_{temp})$. Here, $W(t)$ and $V(t)$ are operators in the 
Heisenberg picture.

In the energy eigenbasis, the thermal OTOC can be expressed as \cite{hashimoto2017out}
\begin{equation}
C_T =
\frac{1}{Z}
\sum_n e^{-\beta E_n} c_n,
\qquad
c_n \equiv
-\langle n|[W(t),V(0)]^2|n\rangle,
\label{eq2.8}
\end{equation}
where $E_n$ and $|n\rangle$ denote the energy eigenvalues and corresponding 
eigenstates of the time-independent Hamiltonian, respectively, and $Z$ is the partition function. 

Using the 
completeness relation $1=\sum_m |m\rangle\langle m|$, the microcanonical 
OTOC can be written as
\begin{equation}
c_n(t)
=
\sum_m b_{nm}(t)\;b_{nm}^{*}(t),
\quad
b_{nm}(t)
=
-i\langle n|\;[W(t),V(0)]\;|m\rangle.
\label{eq2.9}
\end{equation}

The OTOC is evaluated using (\ref{eq2.8}). We choose the two variables as the position and 
momentum operators, i.e., $W(t) = x(t)$ and $V(0) = p(0)$.The inverse temperature is set to $\beta = 0.5$. The operator $x(t)$ is evolved in time using the following equation,
\begin{equation}
    x(t+\delta t) = U^{(4)\dagger}_\mathrm{cf}(t+\delta t,t)\ x(t)\ U^{(4)}_\mathrm{cf}(t+\delta t, t).
\end{equation}
Here, the propagator $U^{(4)}_{\mathrm{cf}}(t+\delta t,t)$ is obtained using the CFET propagation scheme described in Section~\ref{sec:numerics}.

Fig.~\ref{fig:OTOC_all} shows the temporal 
evolution of the OTOC for (a) $\omega_r = 2$, (b) $1.6$, and (c) $(\sqrt{5}+1)/2$, respectively. Each figure shows the full temporal evolution, along with the
early-time behavior on a semi-logarithmic scale in the inset, to identify the exponential growth regime.

For all three forcing-frequency ratios, the OTOC exhibits an initial 
exponential growth with an almost identical scaling behavior at early times. Specifically, the exponent $\lambda \approx 0.33$--$0.34$ is consistent across all three cases. This near-universality of the early-time behavior indicates that the onset of local quantum chaos and short-time scrambling are largely insensitive to the rationality of the driving-frequency ratio.

The extracted values $\lambda \approx 0.33\!-\!0.34$ also satisfy the 
Maldacena-Shenker-Stanford (MSS) bound \cite{maldacena2016bound},
$\lambda \leq 2\pi k_B T_{temp}/\hbar$, which in our units ($\hbar = k_B = 1$) 
and in $\beta = 0.5$ ($T_{temp} = 2$) gives $4\pi \approx 12.57$ thus the ratio 
$\lambda/(2\pi k_BT_{temp}/\hbar) \approx 0.026$.

However, significant differences appear at intermediate and long time scales. For the irrational ratio $\omega_r = (\sqrt{5}+1)/2$ (see Fig.~\ref{fig:OTOC_all}(c)), the OTOC reaches saturation earlier than in the case of $\omega_r = 1.6$ (see Fig.~\ref{fig:OTOC_all}(b)). In contrast, the system driven by integer frequency ratio (see Fig.~\ref{fig:OTOC_all}(a) for $\omega_r = 2$) displays a prolonged, increasing OTOC in the intermediate-time regime, without a clear scaling behavior. The yellow points shown in Fig.~\ref{fig:OTOC_all} represent averages over individual oscillations and are included solely as a visual guide to highlight the overall increasing trend of the OTOC. These averaged points are not used for any quantitative analysis.

\subsubsection{Fidelity}
\label{fed}

Fidelity, denoted by $\mathcal{M}(t)$, quantifies the stability of a quantum state under a small perturbation of the system Hamiltonian. It measures the overlap between two states evolved from the same initial condition under slightly different Hamiltonians, and is defined as
\begin{equation}
    \mathcal{M}(t) = \left| \braket{\psi_0 \left| e^{i{H}_2 t/\hbar} e^{-i{H}_1 t/\hbar} \right| \psi_0} \right|^2 ,
    \label{eq_fid}
\end{equation}
where $\ket{\psi_0}$ is the initial state of the system, ${H}_1$ is the unperturbed Hamiltonian, and the perturbed Hamiltonian is given by
\begin{equation}
    {H}_2 = {H}_1 + \varepsilon V, \label{ev}
\end{equation}
where $\varepsilon$ is the perturbation strength and $V$ is the perturbation operator. 

In the present work, we choose the perturbation operator to be the static part of the Hamiltonian, i.e., $V = {H}_0$. For a time-dependent Hamiltonian, the unitary evolution operators $e^{-i{H}t/\hbar}$ are replaced by the corresponding time-evolution propagators $U(t_2,t_1)$.

An ensemble of five initial conditions is constructed by first
expanding the Gaussian wave packet in the eigen basis of $H_0$. The remaining four members are generated by adding a small random complex perturbation of fixed magnitude $\delta = 0.001$ to the Gaussian wave packet, with each resulting state renormalized before use. We emphasize that this perturbation of the initial state is distinct from the Hamiltonian perturbation $\varepsilon V$ introduced in (\ref{ev}), where $V = H_0$. There, the perturbation strength $\varepsilon$ controls the difference between the two Hamiltonians $H_1$ and $H_2$ under which a single initial state evolves. Here, by contrast, the perturbation is applied to the initial state itself, purely to generate an ensemble of five initial conditions over which the fidelity $\mathcal{M}(t)$ is averaged for statistical robustness. The two perturbations serve different purposes and do not interact. Averaging fidelity over an ensemble of initial states, specifically an ensemble of coherent states, is an established practice to characterize the trend of fidelity decay \cite{gorin2006dynamics}.

\begin{figure}[tbh]
    \centering
    \includegraphics[width=0.9\linewidth]{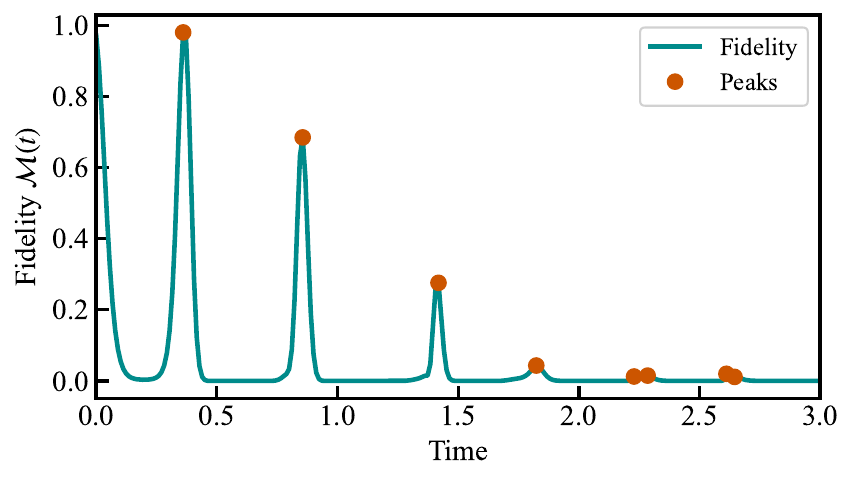}
    \caption{Time evolution of the ensemble-averaged fidelity $\mathcal{M}(t)$, illustrating the identification of successive revival peaks (markers) used to construct the peak-fidelity trend shown in subsequent figures.}
    \label{fid_trend}
\end{figure}

Figure~\ref{fid_trend} illustrates the procedure used to characterize the fidelity decay: $\mathcal{M}(t)$, averaged over an ensemble of $5$ realizations of the initial condition, exhibits a sequence of revivals whose peak amplitude decreases with time. We identify the local maxima of $\mathcal{M}(t)$ and track their peak values, $\mathcal{M}_\text{peak}(t)$, as a compact representation of the decay trend; peaks are extracted using the same procedure for every combination of $\omega_r$ and $\varepsilon$ considered in this work.

\begin{figure*}[tb]
    \centering
    \begin{subfigure}[t]{0.48\textwidth}
        \centering
        \includegraphics[width=\linewidth]{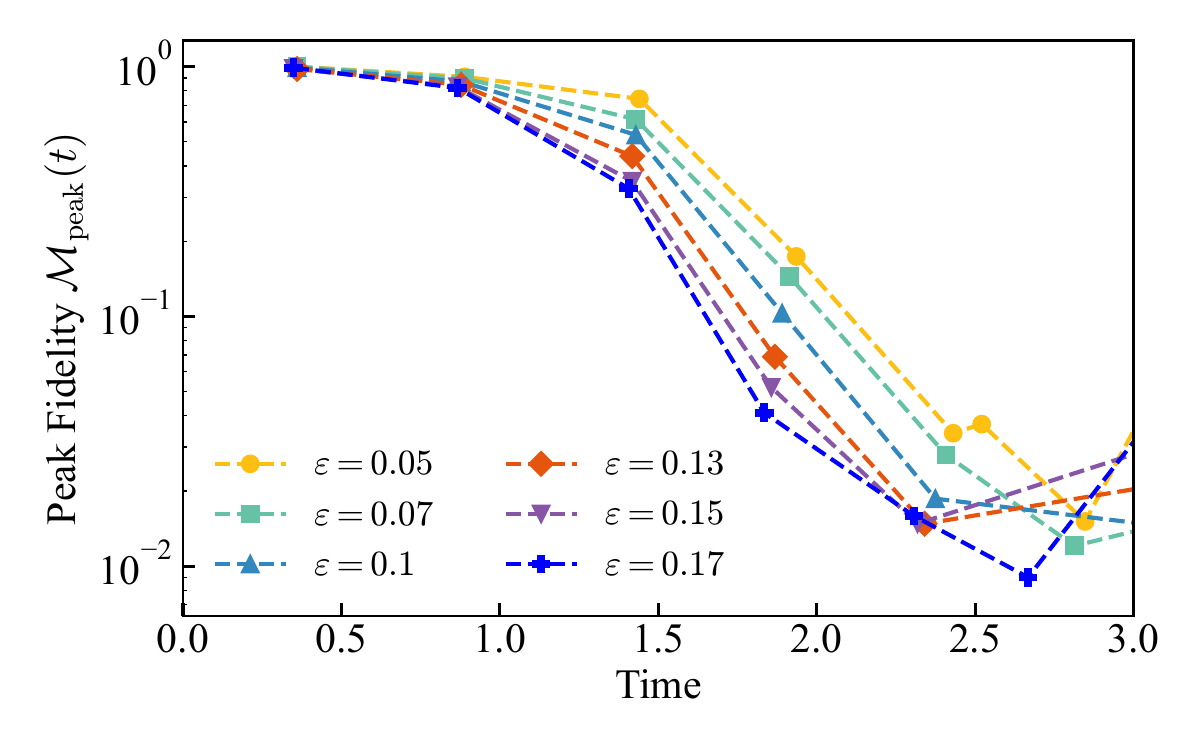}
        \caption{$\omega_r = 2$}
        \label{fidelity:a}
    \end{subfigure}
    \hfill
    \begin{subfigure}[t]{0.48\textwidth}
        \centering
        \includegraphics[width=\linewidth]{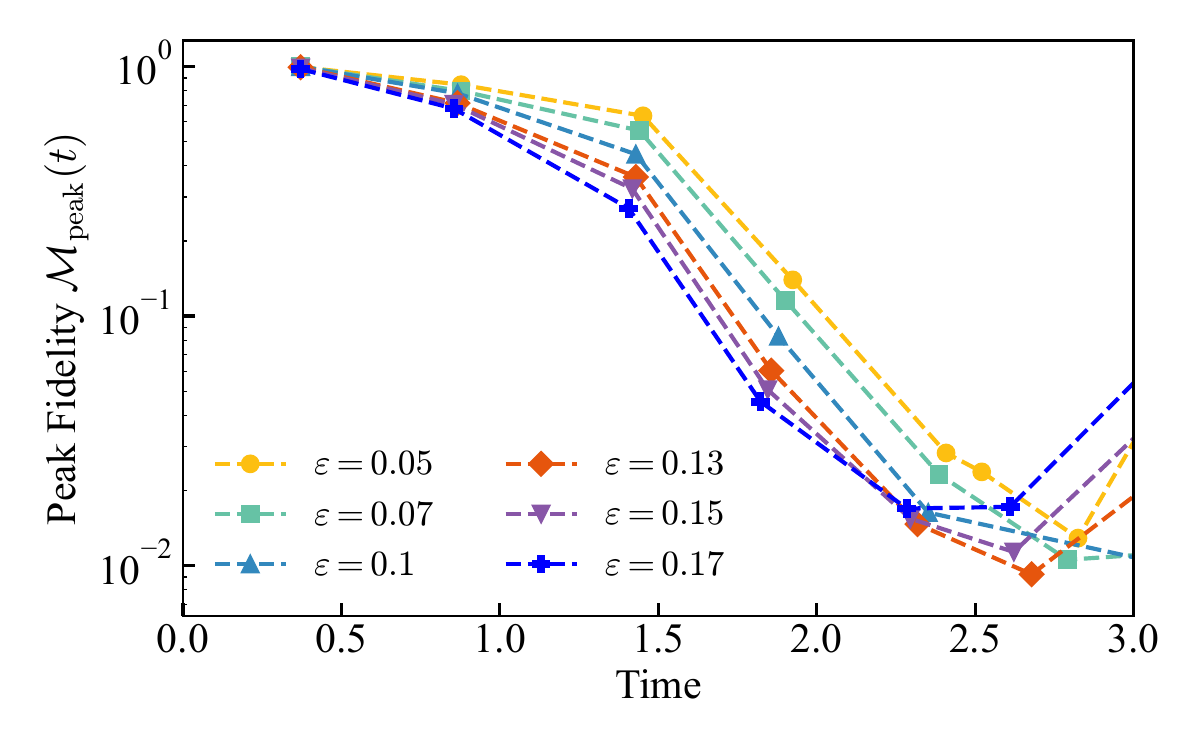}
        \caption{$\omega_r = 1.6$}
        \label{fidelity:b}
    \end{subfigure}

    \vspace{0.8em}

    \begin{subfigure}[t]{0.48\textwidth}
        \centering
        \includegraphics[width=\linewidth]{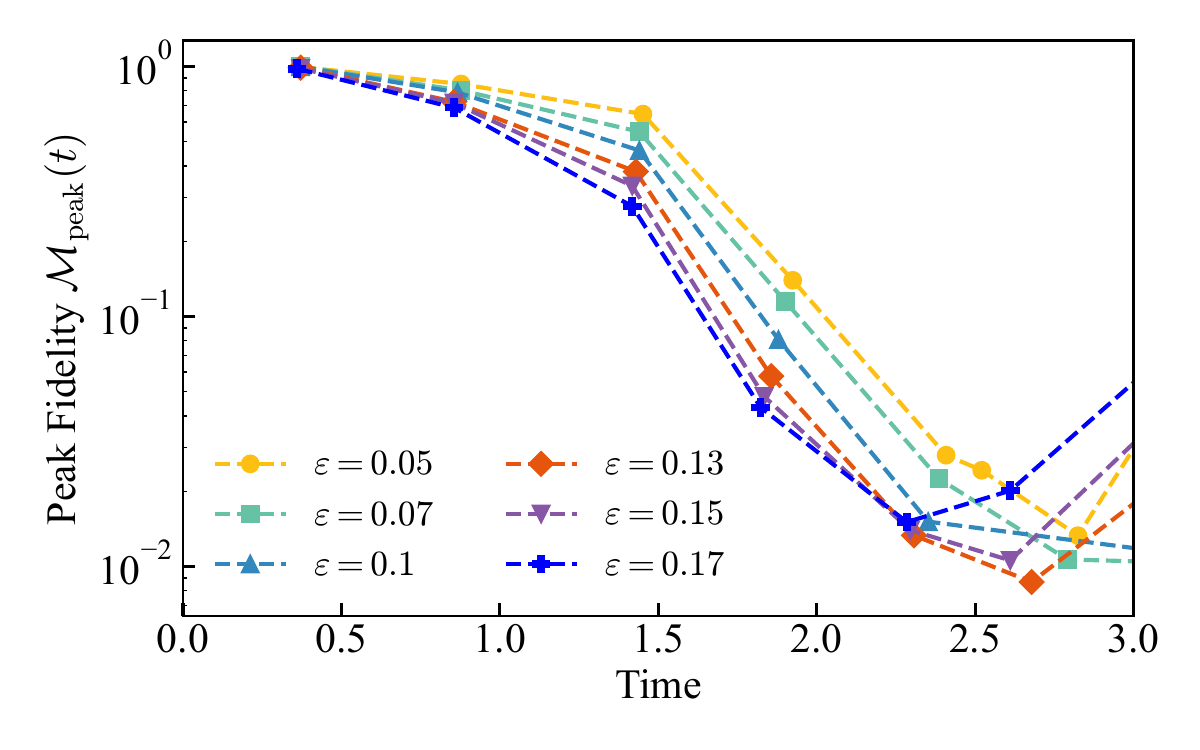}
        \caption{$\omega_r = (\sqrt{5}+1)/2$}
        \label{fidelity:c}
    \end{subfigure}

    \caption{Peak fidelity $\mathcal{M}_\text{peak}(t)$ as a function of time
    for six perturbation strengths $\varepsilon$, shown for three
    forcing-frequency ratios: (a) $\omega_r = 2$, (b) $\omega_r = 1.6$, and
    (c) $\omega_r = (\sqrt{5}+1)/2$.}
    \label{fidelity}
\end{figure*}

Figure~\ref{fidelity} shows the resulting peak-fidelity trend $\mathcal{M}_\text{peak}(t)$ for six perturbation strengths, $\varepsilon = 0.05, 0.07, 0.1, 0.13, 0.15,$ and $0.17$, for three forcing-frequency ratios: $\omega_r = 2$ (Fig.~\ref{fidelity:a}), $\omega_r = 1.6$ (Fig.~\ref{fidelity:b}), and $\omega_r = (\sqrt{5}+1)/2$ (Fig.~\ref{fidelity:c}). For all three frequency ratios, and for every perturbation strength studied, the peak-fidelity trend exhibits the same characteristic three-region behavior. At early times ($t \lesssim 1.5$), $\mathcal{M}_\text{peak}(t)$ decreases relatively slowly, consistent with the short-time, perturbative (quadratic-in-time) decay expected for the fidelity at small $t$ \citep{prosen2002general,gorin2006dynamics}. This is followed by an intermediate window ($1.5 \lesssim t \lesssim 2.3$) in which $\mathcal{M}_\text{peak}(t)$ falls approximately linearly on the semi-logarithmic scale, i.e., an exponential decay of the peak fidelity. At the longest times sampled ($t \gtrsim 2.3$), the peak fidelity reaches a low plateau ($\mathcal{M}_\text{peak} \sim 10^{-2}$) and displays partial revivals rather than continuing to decrease monotonically.

Within the exponential window, the decay curves for the different perturbation strengths are comparable across the range $\varepsilon \in [0.1, 0.17]$ studied here. Since the decay rate is approximately independent of the perturbation strength, this behavior is consistent with a Lyapunov-type regime, in which the loss of fidelity is governed primarily by the intrinsic chaotic dynamics of the system rather than by $\varepsilon$ itself \citep{jalabert2001environment, cucchietti2004universality}. We identify this behavior qualitatively from the decay envelopes in Fig.~\ref{fidelity}. Notably, this signature is observed consistently across all three frequency ratios studied, indicating that the same qualitative decay structure and the same approximate $\varepsilon$-independence of the decay rate in the exponential regime.

\begin{table*}[t]
\centering
\caption{Summary of quantum chaos diagnostics for three forcing-frequency ratios $\omega_r = \omega_1/\omega_2$. FFT,
$K_{\mathrm{median}}$: 0--1 test statistic, FTLE, $\lambda$: early-time OTOC exponential growth rate, OTOC saturation, and  Fidelity trend of the ensemble-averaged peak fidelity $\mathcal{M}_{\mathrm{peak}}(t)$ across perturbation strengths $\varepsilon = 0.05$-$0.17$.}
\label{tab:chaos_summary}
\renewcommand{\arraystretch}{1.4}
\begin{tabular}{cccccccc}
\hline\hline
\multirow{2}{*}{$\omega_r$} & 
\multirow{2}{*}{\makecell{FFT \\ Spectrum}} & 
\multirow{2}{*}{\makecell{$K_{\mathrm{median}}$ \\ (0--1 test)}} & 
\multirow{2}{*}{\makecell{FTLE \\ Distribution}} & 
\multirow{2}{*}{\makecell{OTOC \\ $\lambda$}} & 
\multirow{2}{*}{\makecell{OTOC \\ Saturation}} &
\multirow{2}{*}{\makecell{Fidelity \\ Rate vs.\ $\varepsilon$}} \\
 & & & & & & & \\
\hline
$2$ &  Broadband & $0.73$ & Predominantly positive & $0.343$  & Slow & $\varepsilon$-independent region\\
$1.6$ &  Broadband & $0.84$ & Predominantly positive & $0.335$ & Intermediate & $\varepsilon$-independent region\\
$ ({\sqrt{5}+1})/{2}$ &  Broadband & $0.91$ & Predominantly positive & $0.334$ &  Fast  & $\varepsilon$-independent region \\
\hline\hline
\end{tabular}
\end{table*}

\section{Summary and conclusions}
\label{summary}

In this work, we investigated the dynamics of a quasiperiodically driven impact oscillator, both in its classical and quantum realizations, with particular emphasis on the grazing regime.

For the classical system, varying the wall position $x_w$ revealed a clear transition from quasiperiodic motion to chaos at the grazing boundary, as evidenced by the bifurcation structure, positive Lyapunov exponents, broadband Fourier spectra, and the 0--1 test for chaos.

For the quantum system, the Shannon entropy of the probability density showed broadband spectra, 0--1 test values close to unity, and predominantly positive finite-time Lyapunov exponents for all driving-frequency ratios, providing strong evidence of chaotic dynamics near grazing. OTOC exhibited a common early-time exponential growth rate in all driving protocols, with a quasiperiodic driving frequency $\omega_r = (\sqrt{5}+1)/2$ producing a faster approach to saturation than other frequency ratios. These observations suggest that, while the onset of local chaotic behavior 
and the associated early-time scrambling are not dependent on the irrationality of the frequency ratio, the subsequent route to global scrambling strongly depends on the rationality of the forcing-frequency ratio. The fidelity showed a consistent three-region decay (quadratic, exponential, and saturation with revivals) across all cases, with the exponential-regime decay rate approximately independent of perturbation strength.

Table~\ref{tab:chaos_summary} summarizes the results of all the quantum chaos 
diagnostics used. The consistent agreement across four independent measures---Fourier spectral analysis, the 0--1 test, finite-time Lyapunov exponents, 
and the OTOC---provides strong and mutually corroborating evidence for 
quantum-chaotic behavior near the grazing condition across all 
driving frequency ratios examined.

Together, these results show that quasiperiodic driving induces quantum-chaotic signatures near grazing and enhances scrambling relative to periodic driving. Future work could explore the semiclassical origins of these signatures by examining the system in the $\hbar_{\text{eff} \to 0}$ limit, where the correspondence with the classical grazing transition reported here can be tested directly. It would also be worth incorporating dissipation, for instance, through a quantum Langevin or Lindblad description, to test whether the golden-ratio enhancement of scrambling survives in an open quantum setting. More broadly, a systematic study of the $\hbar_{\text{eff}}$-dependence of these diagnostics would clarify how robust the reported quantum-chaotic signatures are across the crossover from the deep quantum to the semiclassical regime.

\section*{Acknowledgements}
TM acknowledges financial support from UGC, Govt. of India.

\bibliography{ref}

\end{document}